\documentclass[fleqn,usenatbib]{mnras}

\usepackage{newtxtext,newtxmath}

\usepackage[T1]{fontenc}
\usepackage{ae,aecompl}
\usepackage{graphicx}
\graphicspath{{./}{figs/}}

\usepackage{multirow}
\usepackage{amsmath}
\usepackage{url}
\usepackage{hyperref} 
\usepackage{xcolor}

\usepackage{bm}
\hypersetup{
	colorlinks=true,
	linkcolor=red,
	citecolor=blue,
}

\newcommand{\be}{\begin{equation}}
\newcommand{\ee}{\end{equation}}
\newcommand{\bs}{\begin{split}} 
\newcommand{\bea}{\begin{eqnarray}}
\newcommand{\eea}{\end{eqnarray}}
	
\newcommand{\dtbf}{\Delta t_{\rm fit}}
\newcommand{\dtt}{\Delta t_{\rm true}}
\newcommand{\dt}{\Delta t} 
\newcommand{\dti}{\Delta t_i} 
\newcommand{\dttr}{T_{\rm trunc}}
\newcommand{\sigdt}{\sigma_{\Delta t}} 
\newcommand{\nim}{N_{\rm images}}

\title{Deep Learning Unresolved Lensed Lightcurves}

\author[M. Denissenya \& E. V. Linder]{
	Mikhail Denissenya$^1$\thanks{Email: mikhail.denissenya@nu.edu.kz}
	and Eric V.~Linder$^{1,2}$
\\
\\
$^1$ Energetic Cosmos Laboratory, Nazarbayev University, Nur-Sultan 010000, Kazakhstan\\ 
$^2$ Berkeley Center for Cosmological Physics \& Berkeley Lab, 
University of California, Berkeley, CA 94720, USA\\
} 

\begin{document}
\label{firstpage}
\pagerange{\pageref{firstpage}--\pageref{lastpage}}
\maketitle

\begin{abstract}
	Gravitationally lensed sources may have unresolved or blended  
	multiple images, and for time varying sources the lightcurves from 
	individual images can overlap. We use convolutional neural nets 
	to both classify the lightcurves as due to unlensed, double, or 
	quad lensed sources and fit for the time delays. Focusing on lensed supernova systems with time delays $\Delta t\gtrsim6$ 
	days, we achieve 100\% precision and recall in identifying the number of images 
	and then estimating the time delays to $\sigdt\approx1$ day, with a $1000\times$ speedup 
	relative to our previous Monte Carlo technique. This also succeeds for flux noise 
	levels $\sim10\%$. For $\Delta t\in[2,6]$ days we obtain 
	94--98\% accuracy, depending on image configuration. We also explore using partial lightcurves where 
	observations only start near maximum light, without the rise time data, and quantify the success. 
	
\end{abstract}




\section{Introduction} \label{sec:intro} 

Gravitationally lensed Type Ia supernovae should be discovered 
in tens and hundreds in surveys beginning in the next few years 
\citep{2104.05676,1903.00510,1902.05141,2010.12399}. 
These will have intriguing advantages relative to more numerous 
lensed quasars and other sources due to their well defined time 
variation over time scales of months, and standardizable candle nature. 
To increase the numbers of these useful probes of time delay 
cosmology, one might attempt to use lensed systems where the images 
are blended or unresolved (e.g.\ due to lower mass galaxy lenses) 
and the image lightcurves significantly overlap. 

Such overlap poses three basic observational problems: 1) recognition 
of the lightcurve as a lensed Type Ia supernova, when the lightcurve does 
not look like a standard Type Ia supernova, 2) determination of the number 
of images contributing to the lightcurve, and 3) estimation of the 
time delays between all the images. 
One 
part of the first step is identifying the lightcurve as being 
a Type Ia supernova. This is an important initial step. 
Recently developed machine learning tools using spectral information 
accomplish this with $\sim99$\% accuracy and can be potentially 
extended to classify non-supernovae sources as well 
\citep{Davison_2022, Muthukrishna_2019}; also see 
\citet{parsnip} for use of photometric data. Our work here then picks 
up with identifying the Type Ia supernova as a {\it lensed\/} 
Type Ia supernova, and further carries out steps 2 and 3. 

Recent articles have addressed 
these by considering distortions of standard supernova lightcurves, 
by expanding in ``crossing functions'' (basically orthogonal polynomials) with 
arbitrary amplitudes \citep{paper1}, and by free form variation of 
lightcurves with constrained amplitudes \citep{paper2}. 
Both methods mentioned use 
Monte Carlo methods to estimate the time delays, with the second 
method also adding a step to identify robustly the number of images 
in the unresolved systems. Both demonstrated accurate measurement of 
the time delays for $\dt\gtrsim10$ days. In this work, we turn to 
deep learning to accomplish this more quickly, and remove the need 
for separate time delay estimation runs for each potential number of 
images. 
Other work has also investigated various aspects of 
unresolved lightcurves for lensed quasars,  
generally involving longer time delays than we consider, e.g.\ 
\citet{2110.15315,2110.01012,2108.02789,2101.11024,2101.11017,2011.04667}. 
We note that much of our method can be applied to 
more general cases of blended lightcurves, from 
a variety of transients, but we focus here on 
lensed Type Ia supernovae.

In Section~\ref{sec:method} we describe the construction of the 
neural net, training, and test data. Section~\ref{sec:results} 
presents the results for the classification of the number of 
images and time delay estimations, focusing on the previous 
$\dt\gtrsim10$ days range. We investigate 
higher noise systems in 
Sec.~\ref{sec:noise}, 
shorter time delay 
systems in Sec.~\ref{sec:short}, and the use of observational data that misses 
the early time rise in Sec.~\ref{sec:norise}. Discussion of results, 
further work, and conclusions is given in Sec.~\ref{sec:concl}.

\section{Deep Learning Approach} \label{sec:method} 

The basic physical situation is of observation 
of only a single blended lightcurve from 
the combination of unresolved gravitationally lensed multiple images of a time varying source.  
We follow the notation of \citet{paper1,paper2}. The 
observed (blended) flux in a wavelength filter $j$ is 
\be 
F_j(t)=\sum_{i \in \rm images} \mu_i\,U_j(t-\Delta t_i)\,,  
\ee 
where $U(t)$ is the intrinsic, unlensed source flux as a 
function of time (i.e.\ unobservable source lightcurve), 
the sum is over each image $i$ with their individual 
lensing magnifications $\mu_i$ and time delays $\dti$. 
We focus on estimating the observable relative time delays 
$\Delta t_{ij}=\dti-\Delta t_j$ and determining the 
number of images $\nim$. Unlensed systems have 1 image, 
while multiply lensed systems have 2 or 4 images (lensing 
gives an odd number of images but one lensed image is 
generally obscured by the lens galaxy or highly demagnified). 

Unlike \citet{paper1,paper2}, we do not input a form $U(t)$, 
either an expansion about a given form or a free form bounded 
deviation from a given form. Instead we use a training set 
of generated, noisy, lensed and unlensed Type Ia supernova 
(SN Ia) lightcurves and use a convolutional neural net to 
classify the blended lightcurves as arising from an $\nim$ 
system. These are then fed into a convolutional neural net 
to perform the time delay estimation. This same procedure 
could be used for any type of blended lightcurves, 
i.e.\ any lensed transient source, with an appropriate 
training set, though here we focus on SN Ia. 

We explored neural networks without convolution, recurrent neural 
networks, and convolutional neural networks. Convolutional neural 
nets with two hidden convolutional layers yield the best performance; we 
found no significant improvement by including further hidden layers.

\subsection{Training and test data} 

For training, 
we generate three data sets, each containing 10000 systems 
with successive time delays between images in the range $[10,14]$ days, magnification 
ratios in $[0.25,4]$, and 
measurement noise at the level of 5\% of the peak flux 
(so points on the rise and tail of the lightcurve have 
signal to noise much less than 20; also see 
\citet{paper1,paper2} for details), 
using the Hsiao supernova lightcurve template \citep{hsiao} within sncosmo \citep{sncosmo}. Our LCsimulator code used to simulate unlensed and lensed systems is publicly available in a GitHub repository\footnote{LCsimulator \url{https://github.com/mdeatecl/LensedSN124imagesLCs}}. 
Set T124 includes 1-, 2-, and 4-image systems 
in equal proportions, while 
Set T2 includes only 2-image systems and Set T4 includes only 
4-image systems. 
We use Set T124 to train a convolutional neural network we call  
CNNc for  classification (as unlensed, i.e.\ 1-image, 
lensed 2-image, or lensed 4-image systems). 
Data sets 
T2 and T4 train neural nets CNN2 and CNN4 respectively to 
accurately predict the time delays for lensed systems with the indicated number of images. We assign 80\% of the systems for 
training and employ 20\% for testing in each data set.

\subsection{Data preprocessing} 

To improve computational efficiency we take several steps to 
prepare the data for input. 
We normalize the flux measurements and the observation intervals 
using the corresponding ensemble (10000 systems) maximal and 
minimal values. 
We stack flux measurements from the three (g, r, i) 
wavelength filters as well as the observation time instances into 
single input vector.

\subsection{Convolutional Neural Network structure} \label{subsec:cnn} 

\begin{figure*}
	\includegraphics[scale=0.85]{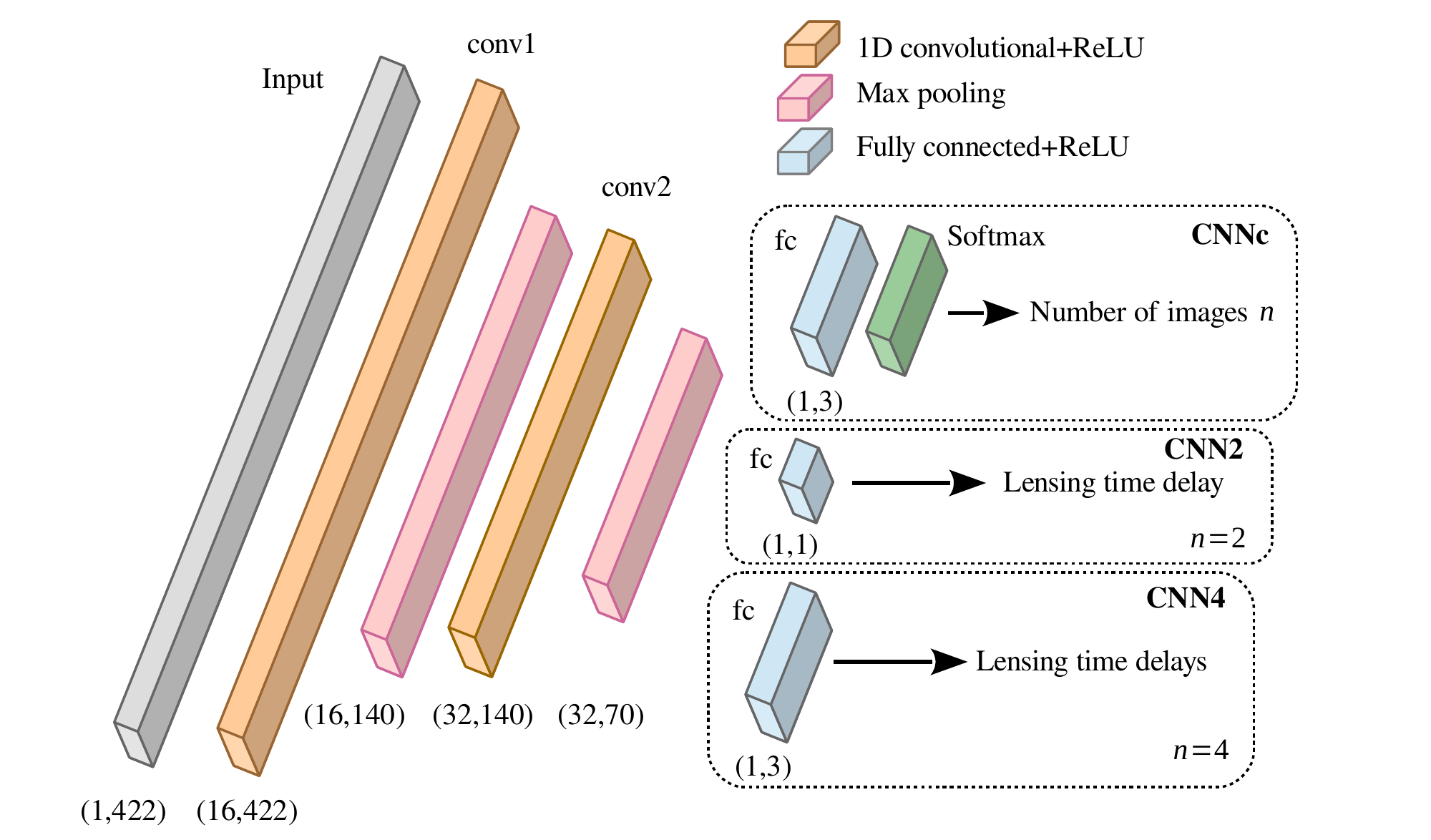}
	\caption{The unified structure of three convolutional neural networks CNNc, CNN2,  and CNN4. The CNNs are designed to have the same combination of inner layers and differ in the output layers indicated enclosed in the dashed rounded boxes.  The maximum probability of the softmax function of the CNNc output determines the number of images $n$ estimated for the system. CNN2 and CNN4 neural networks estimate lensing time delays in the systems with $n=2$ and $n=4$ number of images respectively. 
		CNNc, CNN2, and CNN4 have independent training sets.
	}
	\label{fig:cnn}
\end{figure*}

In this paper, we implement three convolutional neural networks CNNc, CNN2, and CNN4 using the PyTorch high-performance deep learning library \citep{pytorch}. These neural networks consist of five layers described in Fig.~\ref{fig:cnn} and share the same structure of inner layers consisting of alternating convolutional and max pooling layers with ReLU (rectified linear unit) activation functions. The fully connected layers form the outputs of neural networks. The output of CNNc predicts the number of images in the system. Depending on the CNNc output we invoke either CNN2 or CNN4 to estimate lensing time delays. CNN2 has a single output number corresponding to a lensing time delay in a two-image system. If CNNc classifies the system as quadruply imaged we employ CNN4 to estimate the corresponding relative lensing time delays, resulting in the output vector with three entries.  

Each convolutional layer produces multiple copies of an input by convolving it with $K$ kernels.   In our case, the convolution turns the input vector into a tensor with an extra depth index which comes first in a tuple defining the size of the output. The depth of the output is equal to the number of kernels $K$ used to transform the input to output data. 

We have confirmed that assigning the tasks of image number classification and estimation of lensing time delays to separate neural networks as we do is advantageous over a neural network architecture executing these task simultaneously. For example, one could use CNNc including the CNN4 output layer dedicated to estimating the 3 lensing time delays in the 4-image systems. While such a neural network would still be capable of reliably identifying the number of images, the presence of unlensed and 2-image systems in the training sample but fitting for (potentially null) 3 time delays skews the distributions of estimated time delays even for the 4-image systems. Thus we use 
the more robust architecture of CNNc followed by CNN2 or CNN4.

\section{Results} \label{sec:results} 

As described in Sec.~\ref{sec:method} we train the CNNs and then 
validate them on data not used in training. 
For 
data from an actual survey, our goal is to 
productively select systems for follow-up observations with 
high-resolution instruments, to enable use of our time delay 
estimations as a cosmological probe, covering the key range of 
lensing time delays 6--14 days \citep{goldstein}. This 
section assesses the purity of the image classification 
and time delay estimation, and we carry out some studies 
of the efficiency (which will depend on survey characteristics 
beyond this work) in Sec. ~\ref{sec:extend}.
Performance of the 
classification, through CNNc, can be quantified with a confusion 
matrix: for each input class, what fraction is classified in each 
output class. We find 100\% precision and recall in predicting the correct 
number of images in the 2000 
lensing systems in the test set. 

\begin{figure*}
	\includegraphics[width=0.48\textwidth]{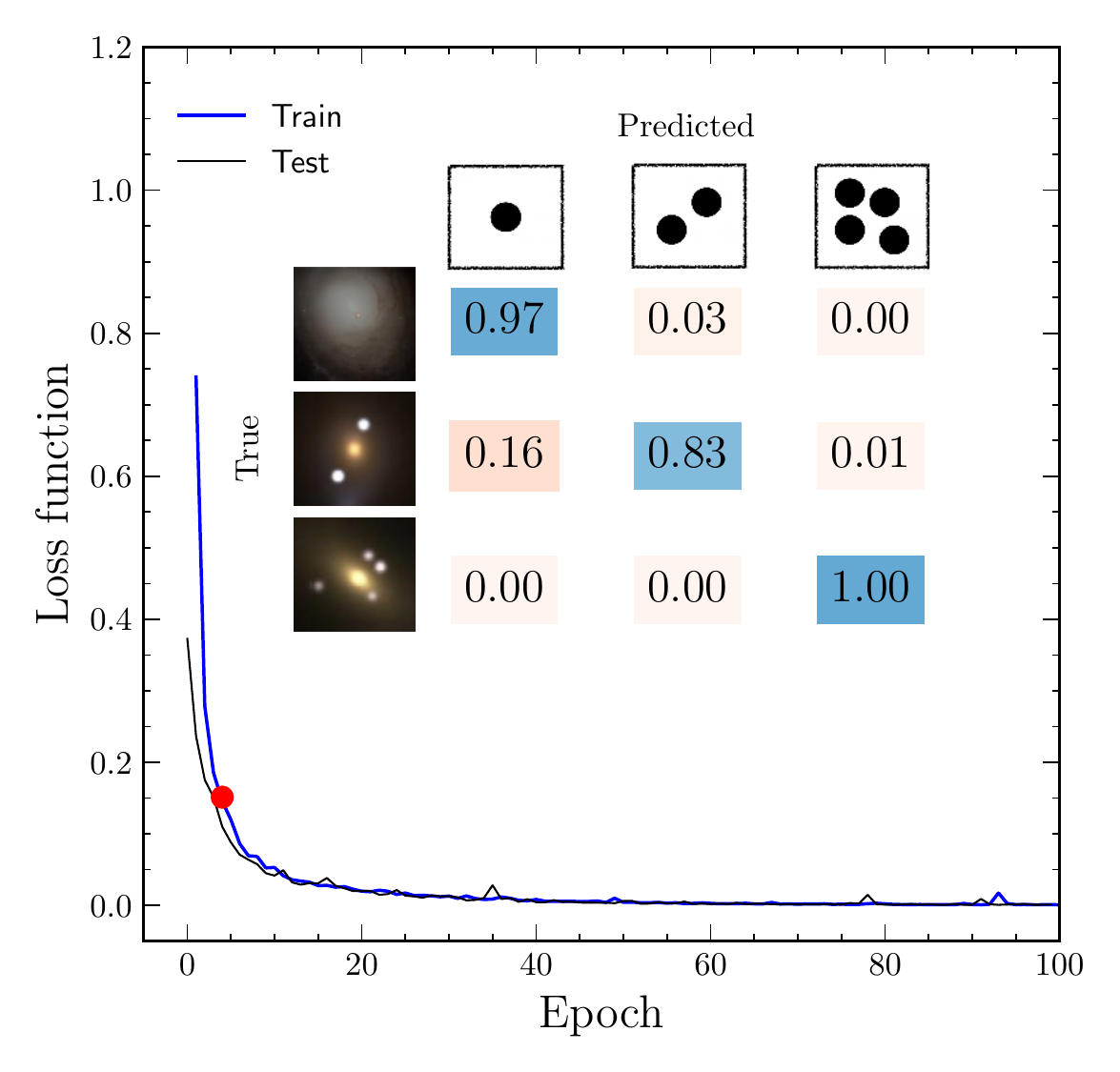}
	\includegraphics[width=0.48\textwidth]{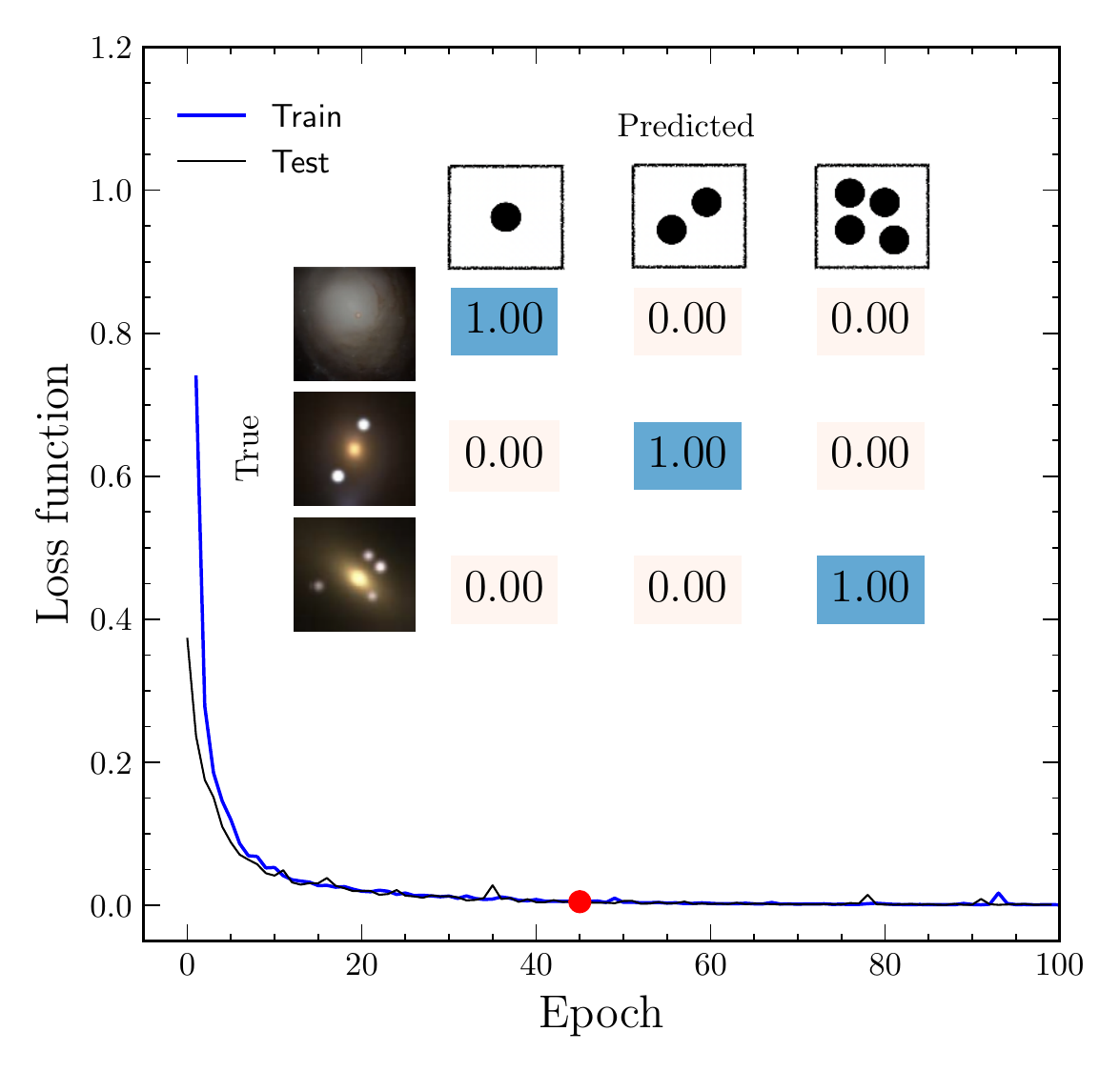}   
	\caption{
		Total loss functions of the neural net CNNc and confusion matrices at  
		epoch 4 (left panel) and 45 (right panel) obtained using the Adam algorithm \citep{kingma2017adam} on a test set. 
		The confusion matrix is diagonalized at epochs above 45 (matrix rows are input classes -- unlensed 1-, 
		lensed 2-, lensed 4-image systems; columns are 
		output classes). 
	}
	\label{fig:confm}
\end{figure*}

Figure~\ref{fig:confm} shows the rapid convergence of CNNc 
toward a perfect, diagonal confusion matrix. There are no false 
assignments. 
The training process takes about $\sim1$ sec for a single epoch (40 iterations), so about 45 sec to 
reach the diagonal confusion matrix shown in the 
figure. Testing of 2000 systems takes  $\sim0.1$ sec on a 4-core CPU operating at 3.5 GHz.

\begin{figure*}
	\includegraphics[width=0.48\textwidth]{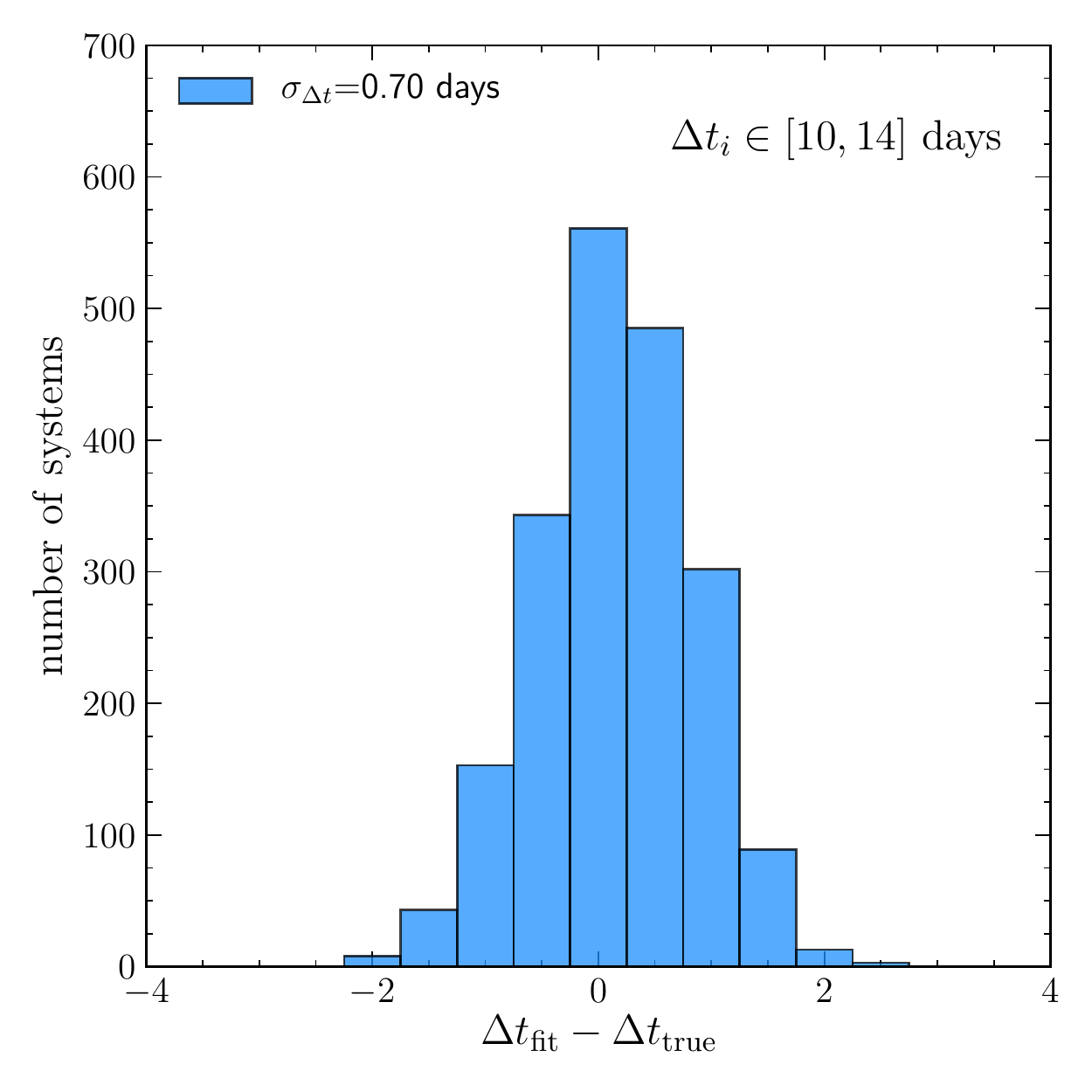}
	\includegraphics[width=0.48\textwidth]{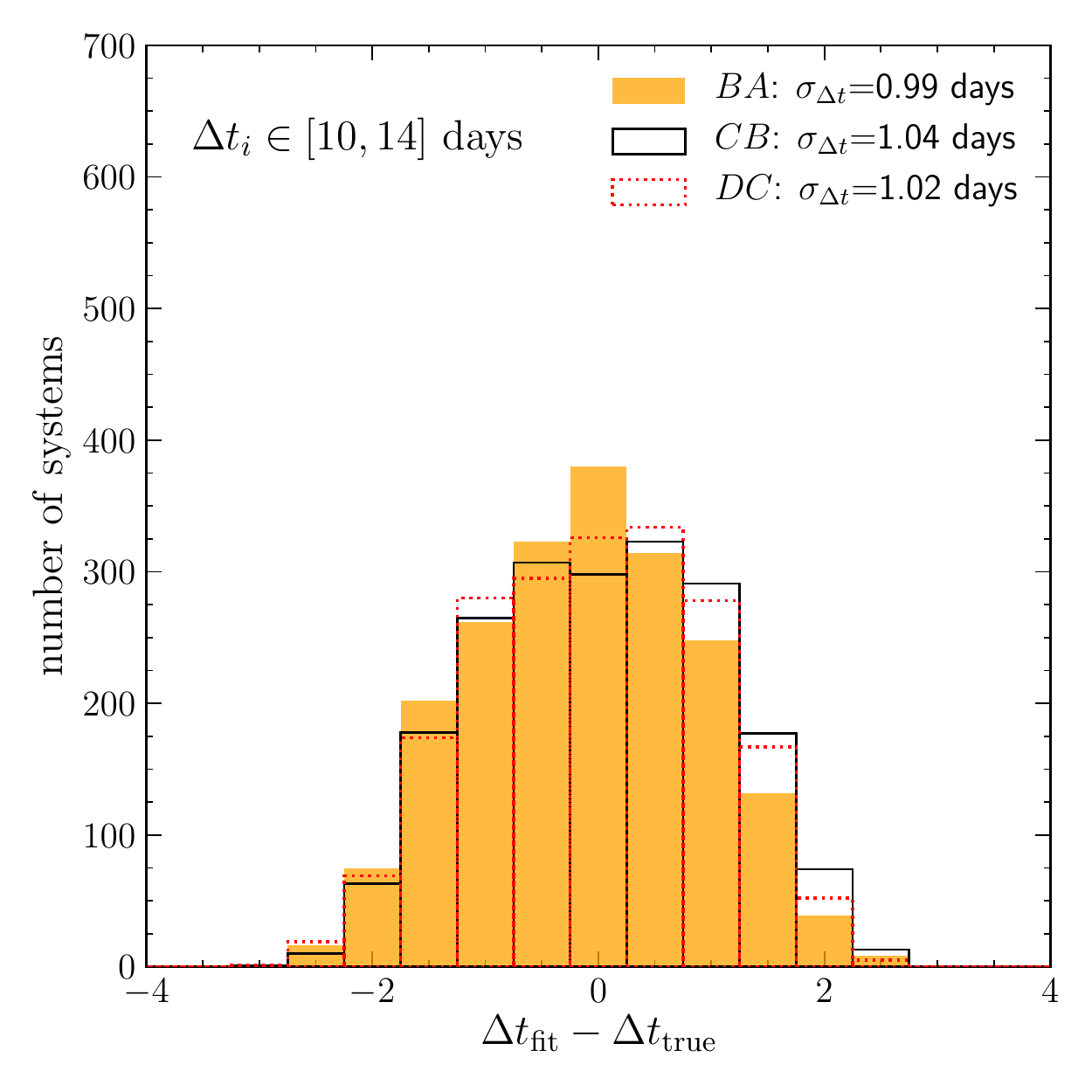}   
	\caption{ 
		Histograms of time delays obtained for the test data sets having $\nim=2$ (left panel) and $\nim=4$ (right panel) using the trained CNN2 and CNN4 neural networks respectively. The left panel 
		histogram shows the single time delay between the two images while 
		the right panel shows the three consecutive time delays between the four 
		images A, B, C, D. The histograms are quantified by the standard deviation  
		$\sigma_{\dt}$, and unbiased to $\lesssim0.4\sigma$. 
	}
	\label{fig:hists5pct}
\end{figure*}

Given the perfect classification, each system is unambiguously 
assigned to either CNN2 or CNN4 for estimation of the  
time delays between the two images A, B, or the four images A, B, C, D. 
Figure~\ref{fig:hists5pct} displays those results as 
a histogram of the offset of the fit value from the true value, 
$\dtbf-\dtt$. The histograms are well peaked and fairly symmetric.

Time delay estimation precision $\sigdt\approx 1$ day for 4-image systems 
and even better for 2-image systems. This is quite satisfactory: 
recall the time delays entering the blended lightcurves are in 
the range $\dt=[10,14]$, and the observing cadence mean is 2 days. (From here on, all time delays are 
understood to be in units of days.) 
The bias, i.e.\ magnitude of systematic offsets from the true 
time delays, is $\lesssim0.4\sigma$, so below the statistical scatter.

The CNN2 and CNN4 
training process takes $\sim1$ sec per epoch (40 iterations);  
the testing of 2000 systems by CNN2 takes $\sim0.1$ sec on a 4-core CPU operating at 3.5 GHz, and testing 2000 systems 
by CNN4 takes roughly the same amount of time.

\section{Extending the Results} \label{sec:extend} 

We next take brief looks at exploring variations, 
one at a time, of the fiducial data 
situation to extend the usefulness of this deep learning 
technique. Section~\ref{sec:noise} assesses 
the impact of data quality by doubling the measurement 
noise. 
In Section~\ref{sec:short} we test the method by decreasing the 
input time delays, first down to 6 days, then all 
the way down to 2 days, giving even greater 
blending of lightcurves. Section~\ref{sec:norise} investigates the 
consequences of missing data from the initial rising 
phase of the observed lightcurve.

\subsection{Noisier Data} \label{sec:noise} 

\begin{figure*}
	\centering
	\includegraphics[width=0.48\textwidth]{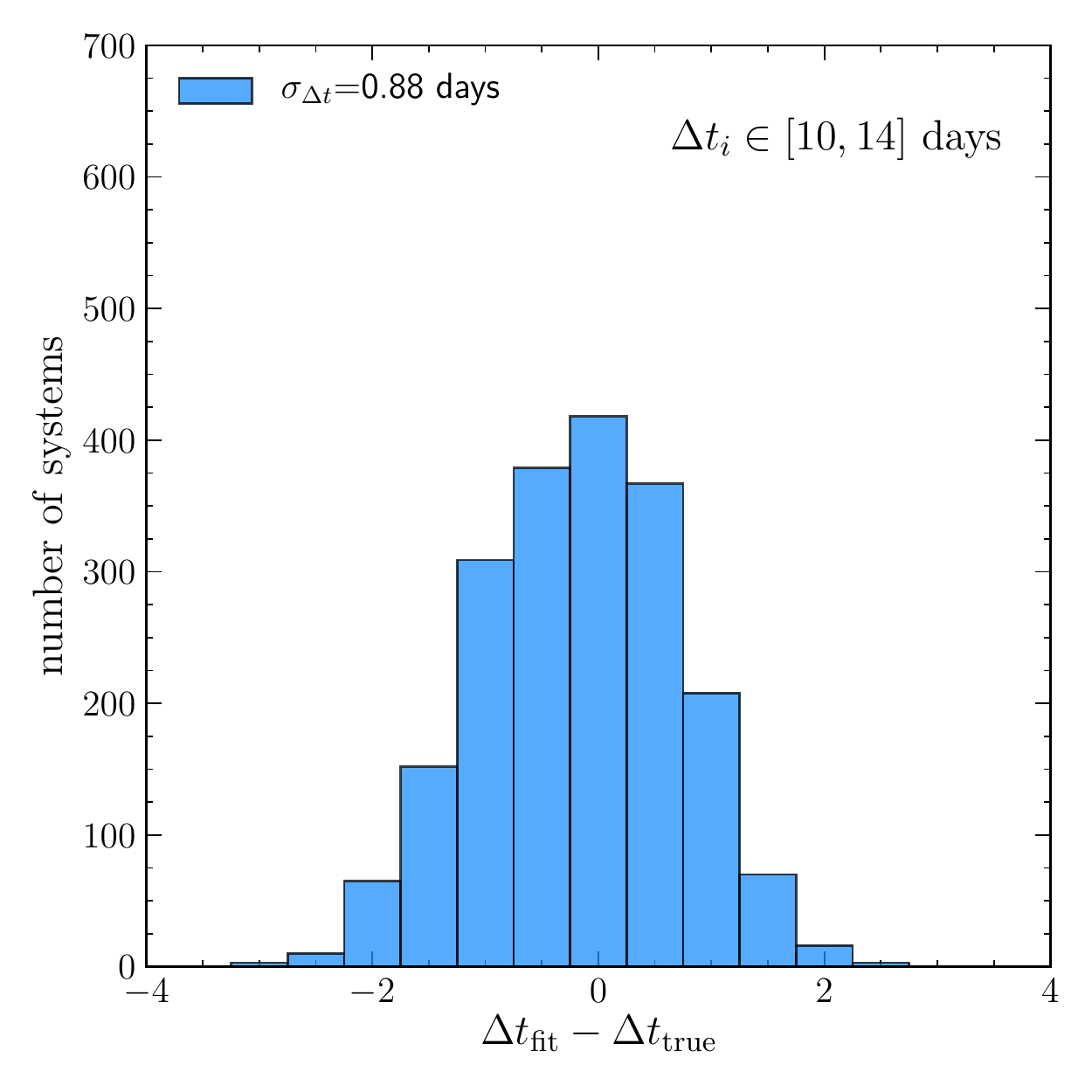}
	\includegraphics[width=0.48\textwidth]{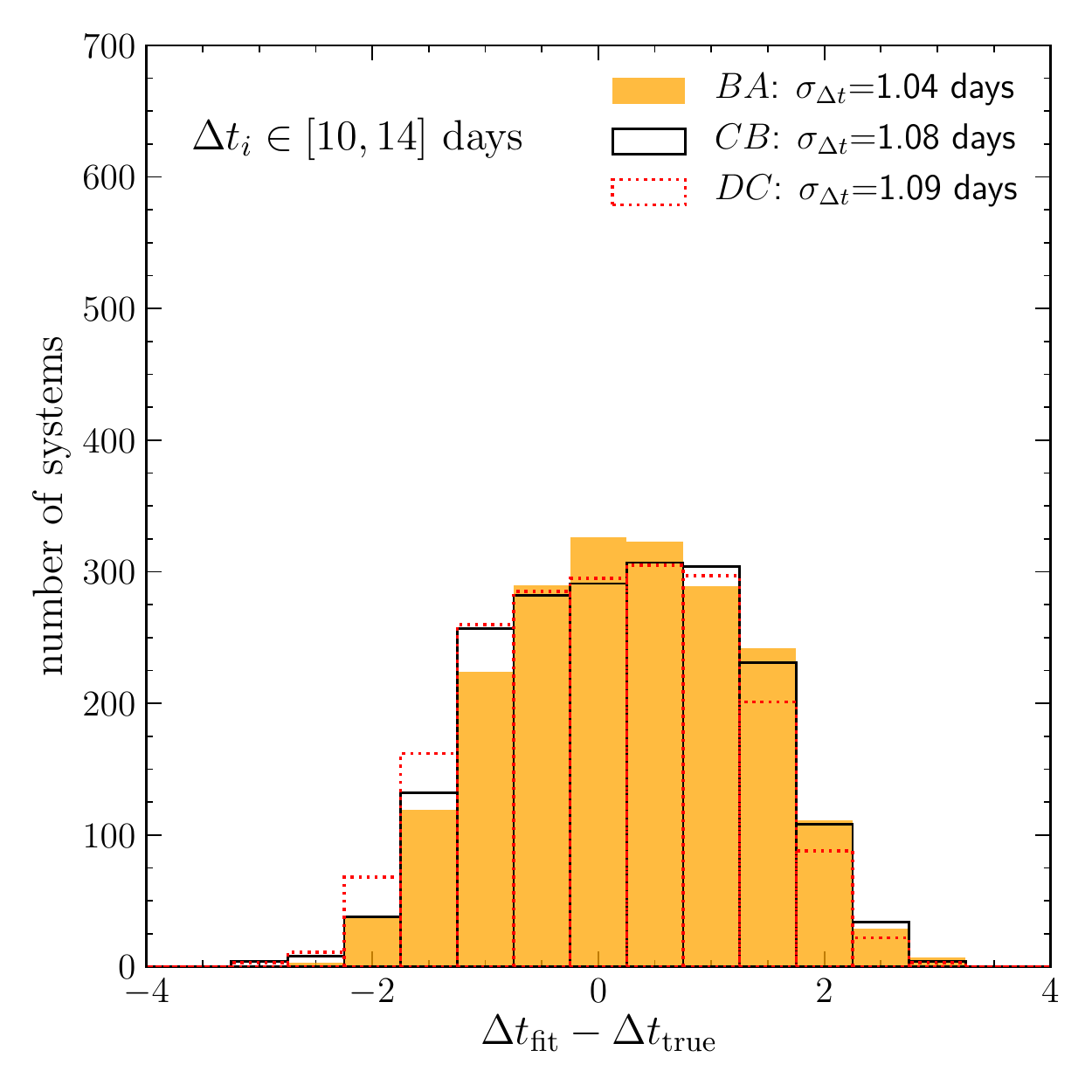}    
	\caption{
		As Fig.~\ref{fig:hists5pct} but for 10\% flux noise. Time delay estimation remains robust. 
	}
	\label{fig:hists10pct}
\end{figure*}

To investigate the impact of measurement noise on the 
classification and time delay estimation, we 
generated another ensemble with flux noise level increased 
to 10\%, but using the same distribution of true time delays and 
magnifications as the fiducial data with 5\% noise level.  
(To allow a direct comparison of the noise impact, we have 
kept the random seeds the same between the 5\% and 10\% sets.) 
We 
found that a perfectly diagonal confusion matrix, i.e.\ 
100\% precision and recall in predicting the number of images, still 
occurs in essentially the same number of iterations of CNNc, despite 
a noise level increased to 10\%. 
Figure~\ref{fig:hists10pct} shows that 
the distribution of predicted time delays has a slightly   
larger spread for the 2-image case compared to the fiducial 
5\% noise case shown in Figure~\ref{fig:hists5pct}, but there 
is little impact on the histogram width in the 4-image case. 
Further studies show that the classification begins to degrade 
at $\sim$20\% noise (much higher than expected from upcoming 
surveys).

\subsection{Shorter Time Delays} \label{sec:short} 

In all other sections we use a fiducial range of 
time delays $\dt=[10,14]$ for comparison to the previous work of \citet{paper1,paper2}. Recall that longer time 
delays would give more clearly resolved lightcurve 
peaks (hence images), while the previous techniques 
began to falter below $\dt=10$. Here we explore how 
the deep learning technique manages on shorter time 
delays. 

For an ensemble of systems with $\dt=[6,10]$ 
generated for training and testing with the same 
neural networks, the CNNc confusion matrix continues to be diagonal over this range $\dt=[6,10]$. 
We have also verified this for the extended range 
$\dt=[6,14]$. Finally, we pushed to even lower 
time delays, $\dt=[2,6]$. Here lensed 2-image 
systems can be mistaken for a single unlensed but 
broader system $\sim5\%$ of the time, while true 
unlensed or 4-image systems are accurately 
classified $\sim98\%$ of the time. 

\begin{figure*}
	\centering
	\includegraphics[width=0.48\textwidth]{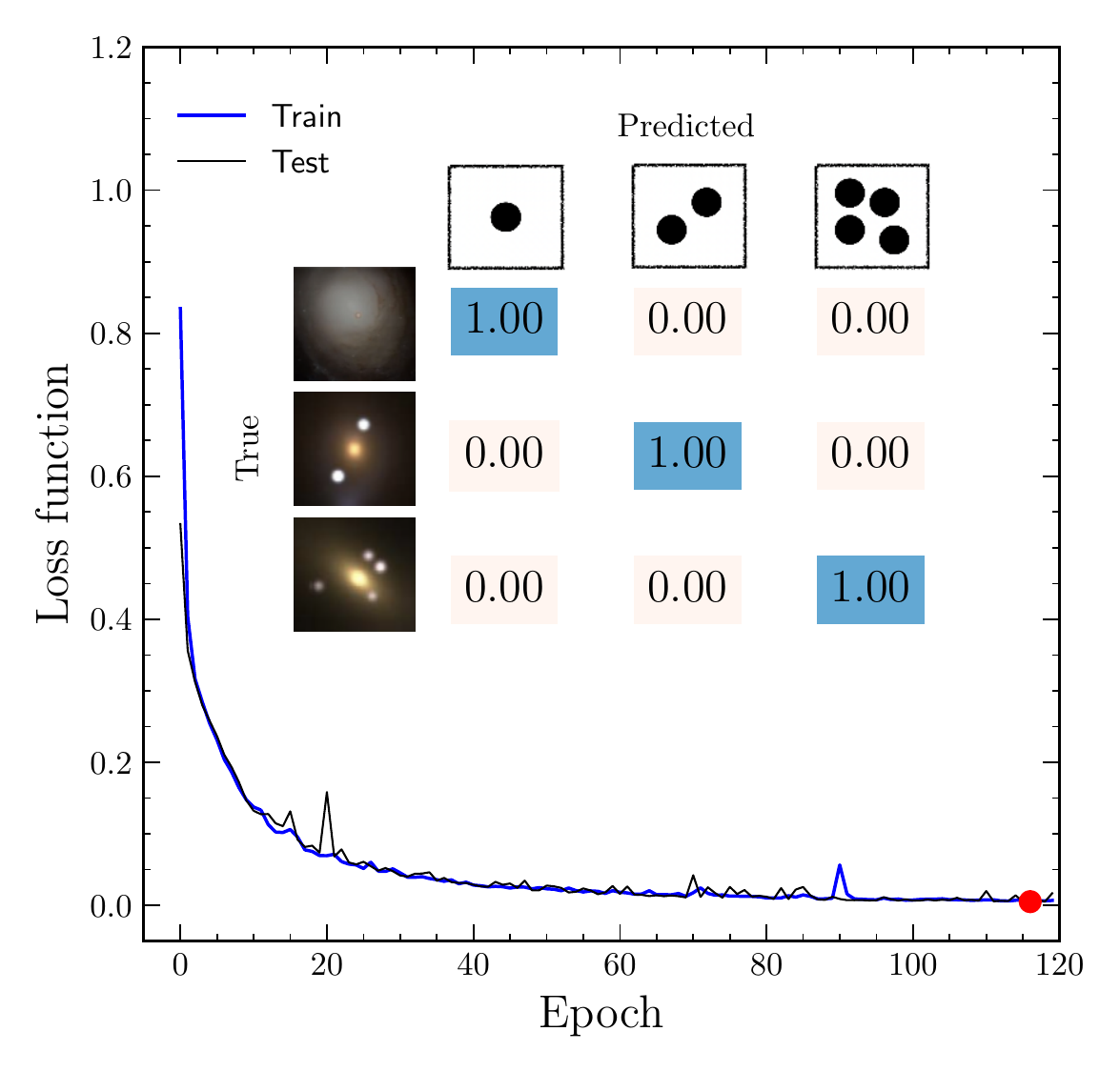}
	\includegraphics[width=0.48\textwidth]{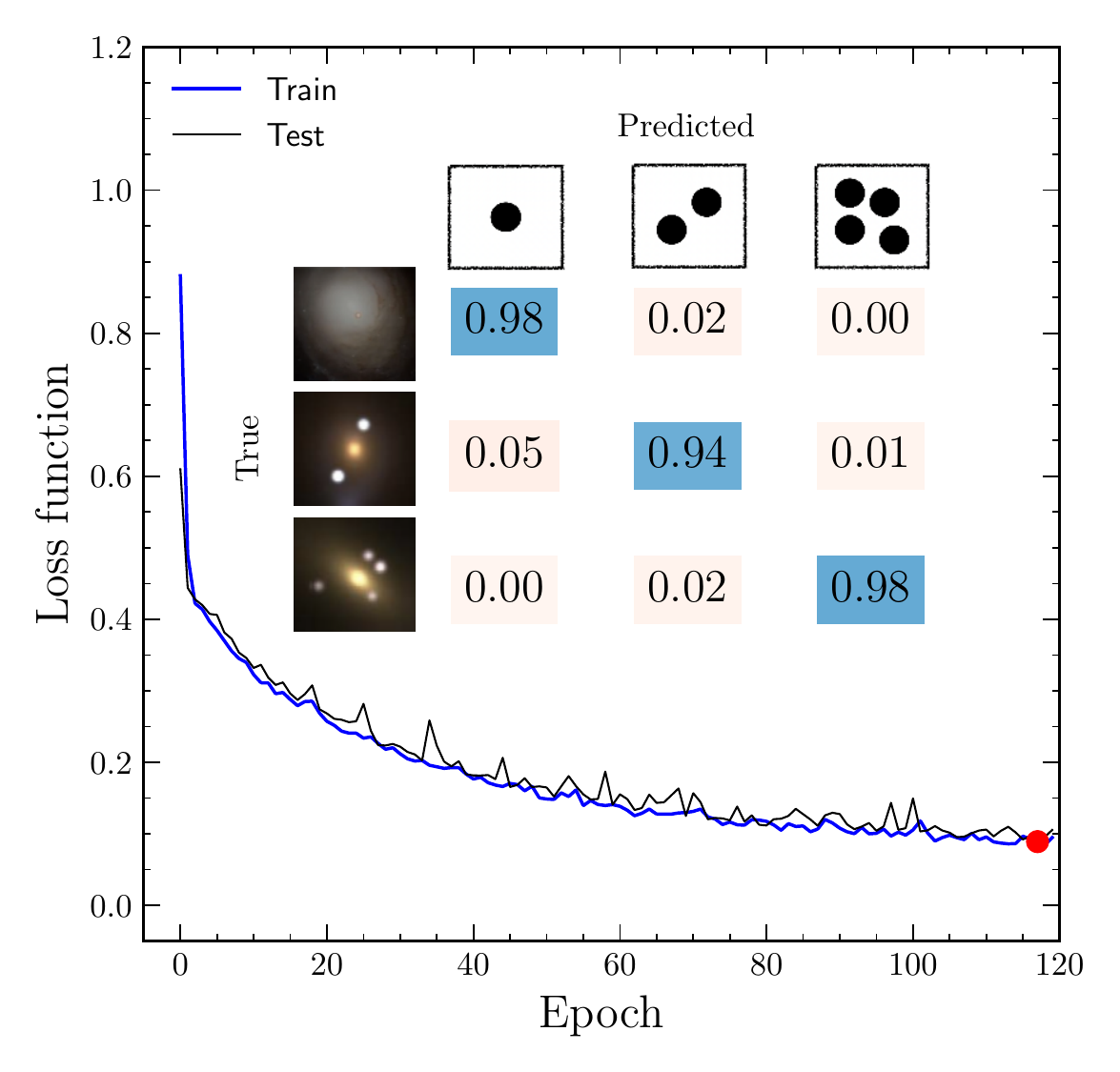} 
	\caption{
		The loss functions of the neural CNNc net  and confusion matrices evaluated for short time delays in the range [6,10] (left panel) and [2,6] (right panel). 
	}
	\label{fig:lowdtl}
\end{figure*}

\begin{figure*}
	\centering
	\includegraphics[width=0.48\textwidth]{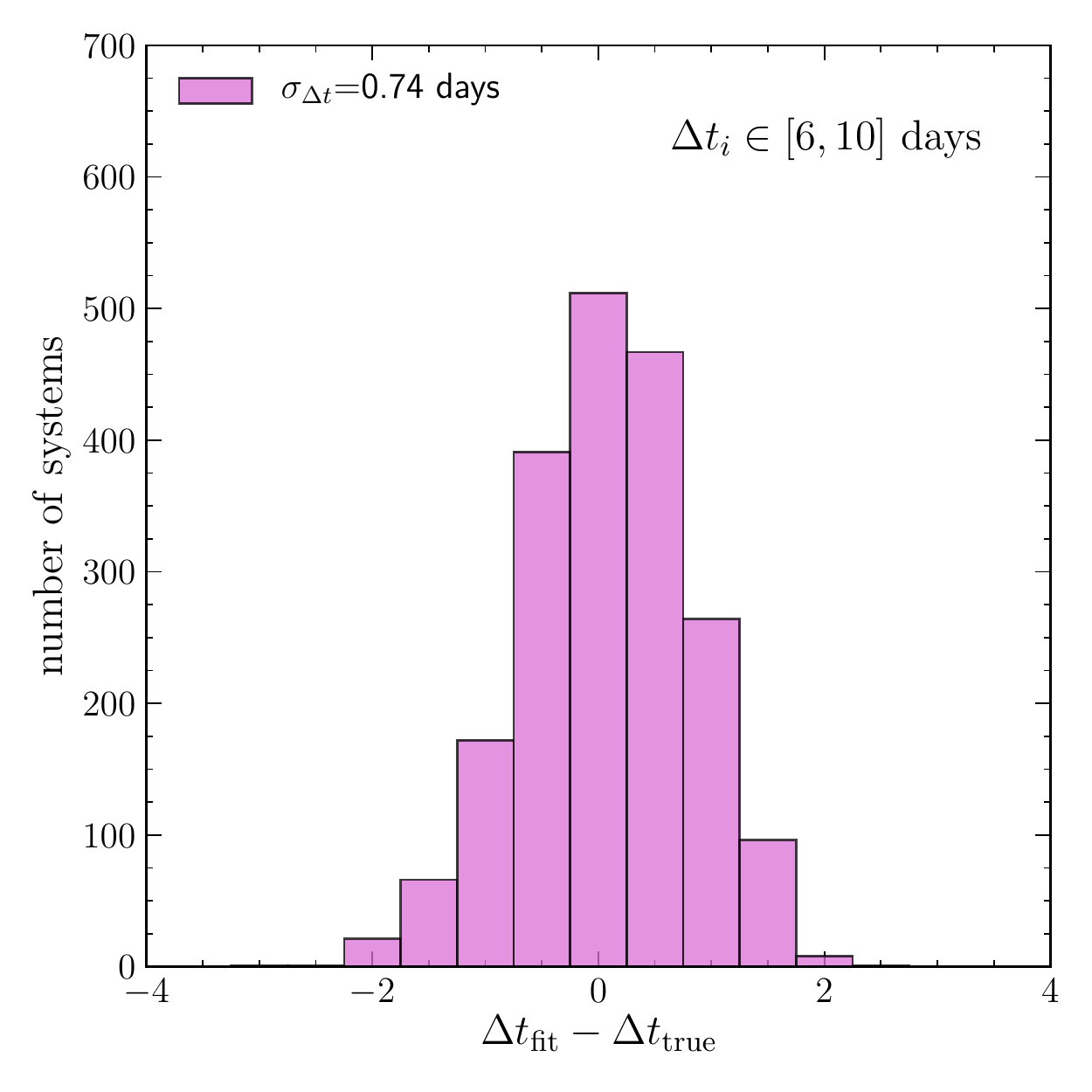}
	\includegraphics[width=0.48\textwidth]{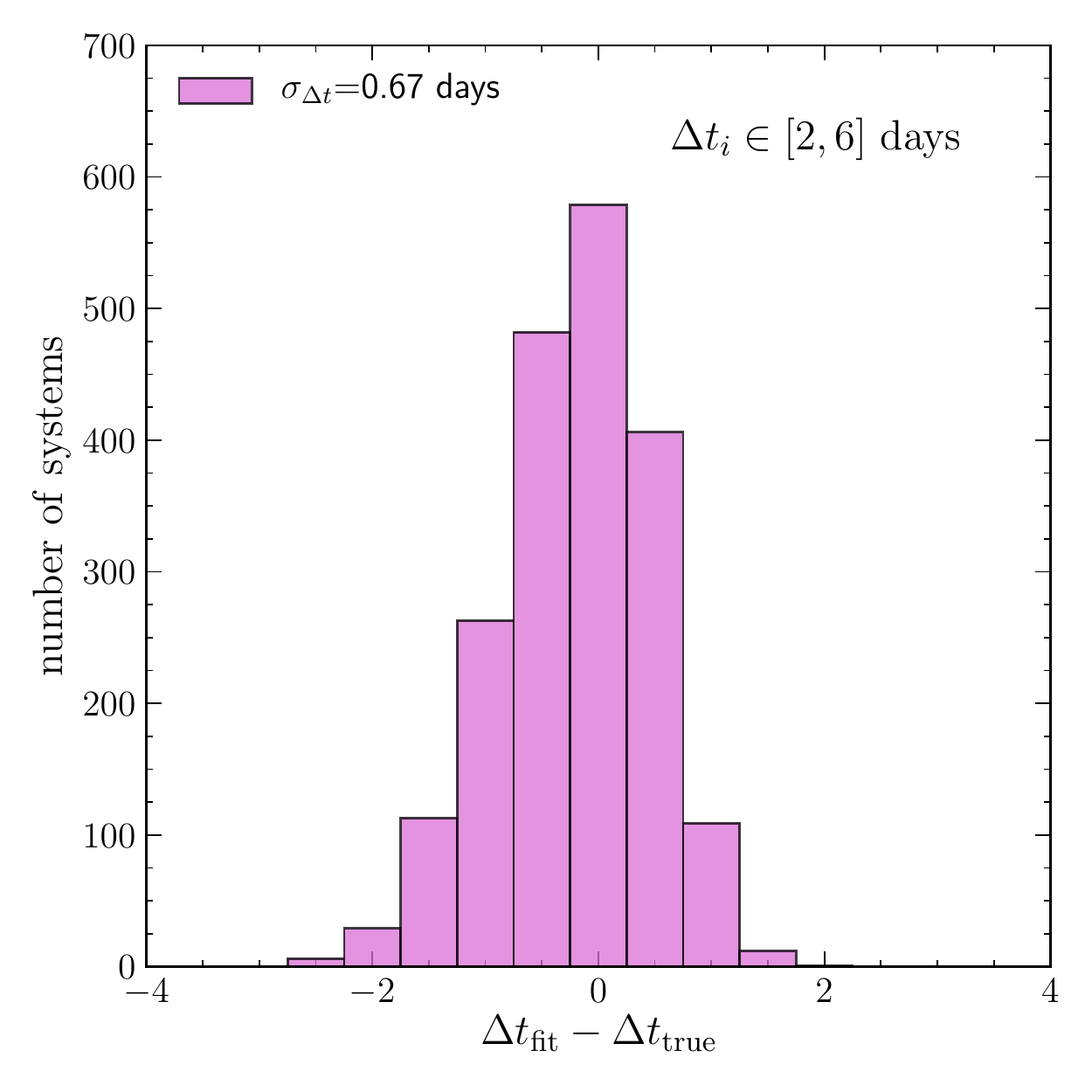} \\
	\includegraphics[width=0.48\textwidth]{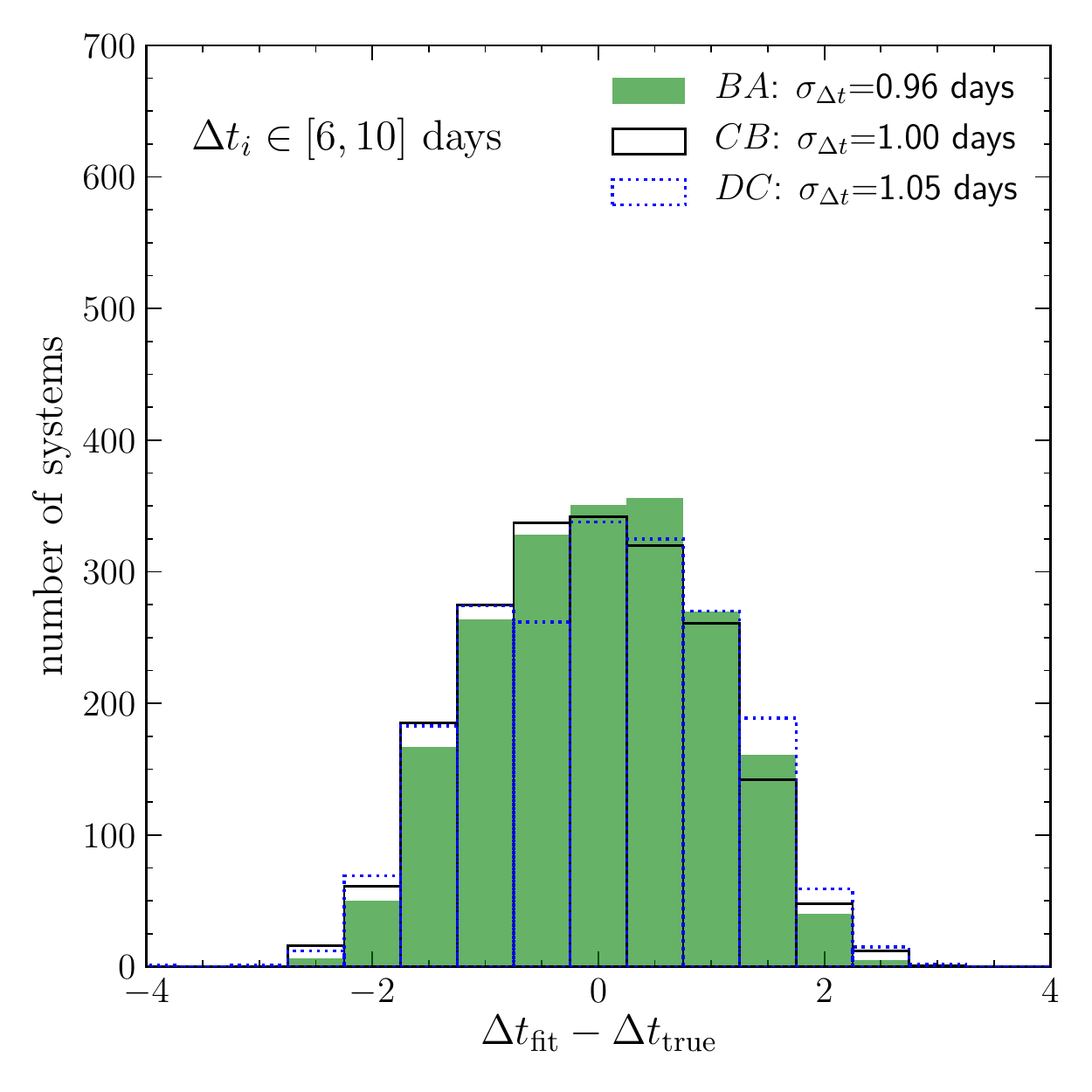}
	\includegraphics[width=0.48\textwidth]{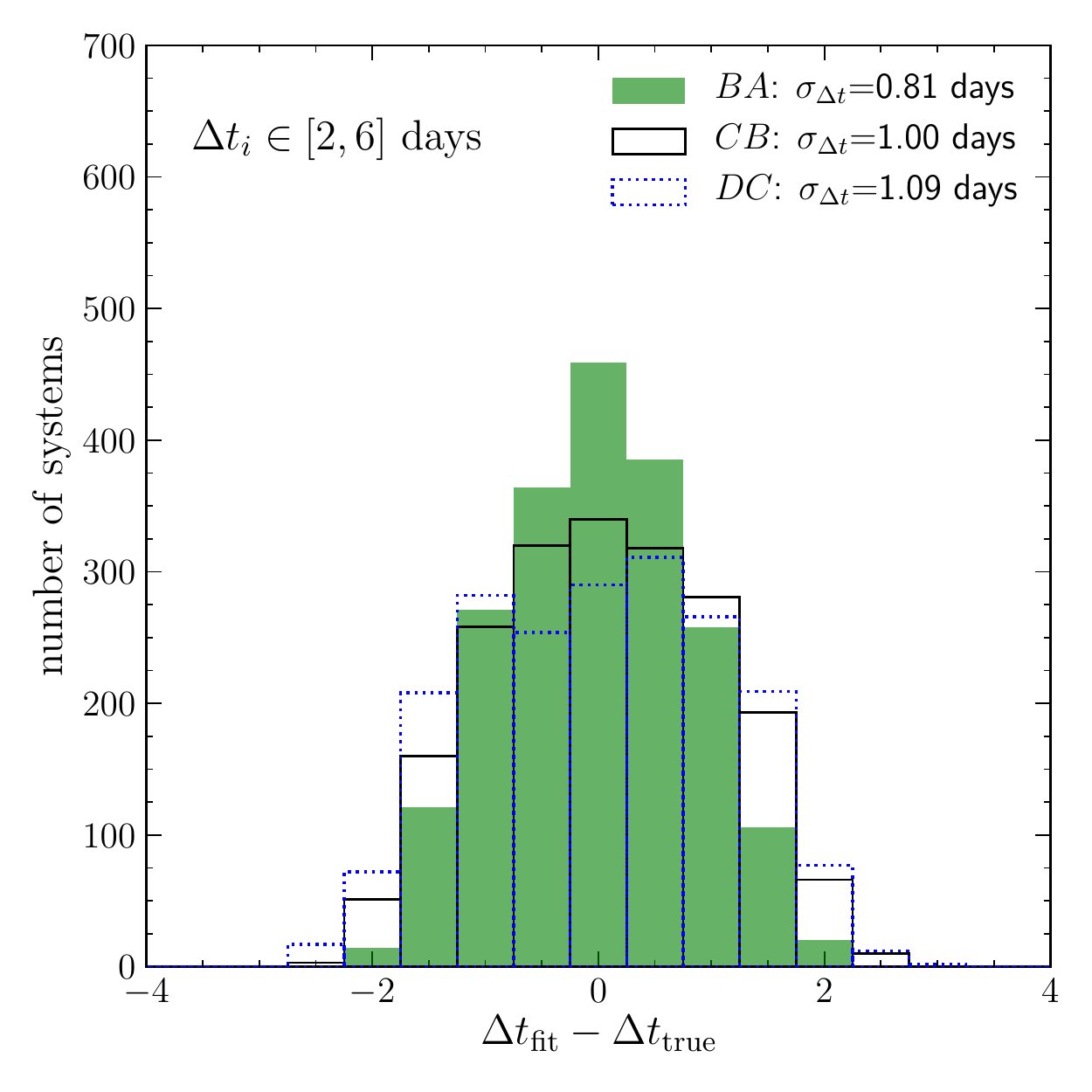} 
	\caption{
		As Fig.~\ref{fig:hists5pct} but for short time delays   in the range [6,10] (left column) and [2,6] (right column). 
	}
	\label{fig:lowdth}
\end{figure*}

Figure~\ref{fig:lowdtl} shows the confusion matrices 
for the $\dt=[6,10]$ and $\dt=[2,6]$ runs. While 
the neural net needs to run nearly three times as long to reach  
the loss function minimum, the end results are, as stated 
above, quite good. Thus, our deep learning technique 
appears robust for $\dt\gtrsim6$. 
Figure~\ref{fig:lowdth} presents the time delay 
estimation histogram for the two low time delay 
ranges, for the classified 2- and 
4-image systems.  
The magnitudes $\sigdt$ do not change 
much, though of course the fractional 
$\sigdt/\dt$ increases going from 
the $\dt=[10,14]$ case to the $[6,10]$ and then $[2,6]$ cases.

\subsection{Missing Early Time Data} \label{sec:norise} 

Observations may not capture the full lensed 
supernova lightcurve data. We investigate the impact on 
classification and time delay estimation 
from missing data in the first $\dttr$ days after 
the supernova explosion. 
We generated multiple training sets of partial lightcurves by truncating the rising part of the full (untruncated) lightcurves. The truncation parameter $\dttr$ indicates the number of days missing after the ideal trigger position determined at 1\% of the total flux peak value for every lightcurve. To investigate the sensitivity of the CNNs to truncation level, we evaluated the performance for several cases: fixed $\dttr= 10, 20, \dots, 80$ for each data set, and the truncation varying in the range $\dttr=[10,40]$. 

\begin{figure*}
	\centering
	\includegraphics[scale=0.85]{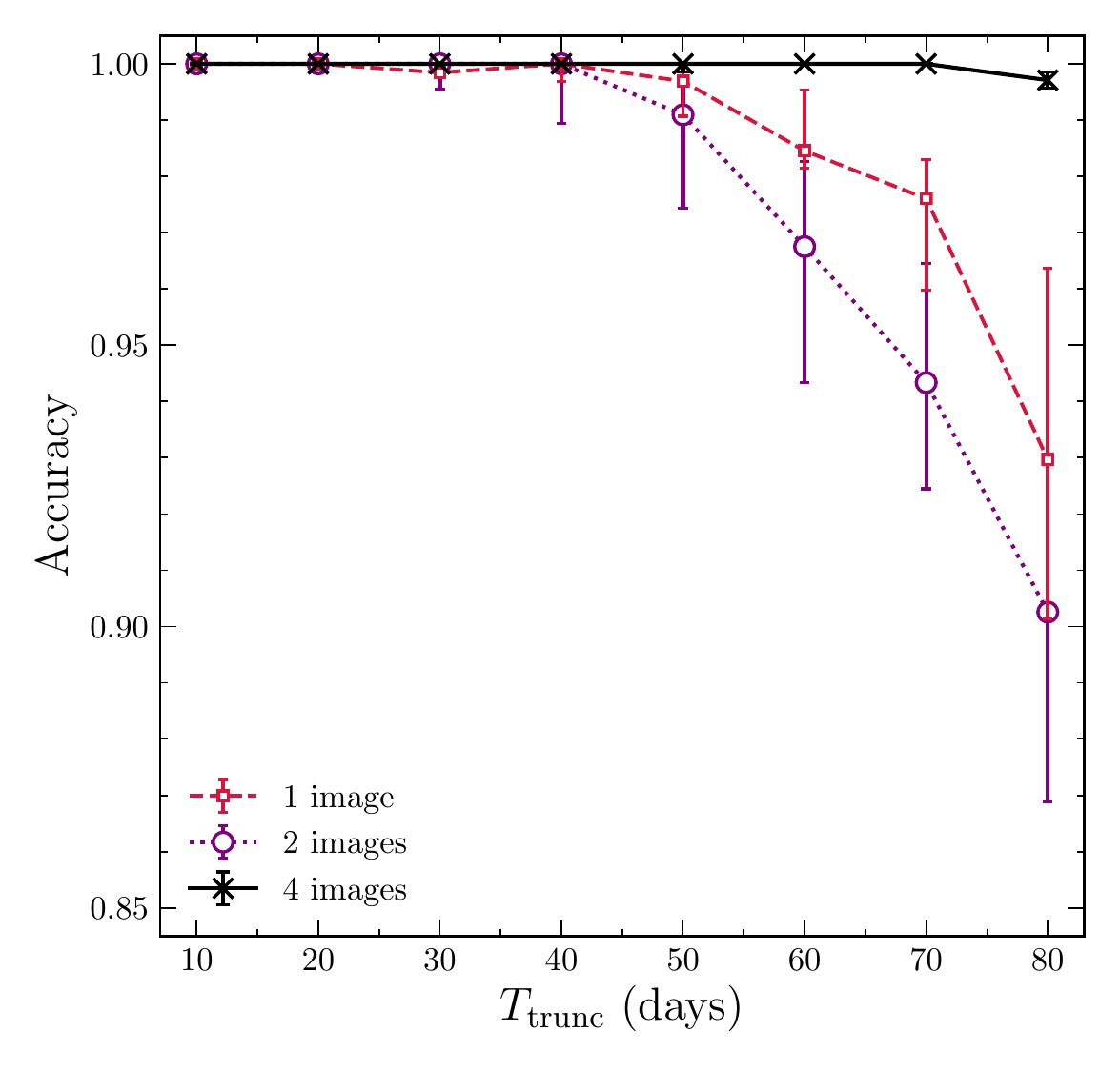}
	\caption{CNNc classification performance computed on the test sets of partial lightcurves at a sequence of data truncation levels. Accuracy refers 
		to the diagonal entries in the confusion matrix. 
	}
	\label{fig:trunc}
\end{figure*}

Figure~\ref{fig:trunc} shows the results as a 
function of $\dttr$ (we find that the case when $\dttr$ varies 
over a range simply falls in between the results 
corresponding to the ends of the range). The recognition of 
unresolved lensed images and classification of 
number of images remains quite robust. Defining 
the accuracy as the diagonal entry in the confusion 
matrix, we find that CNNc delivers accuracies 
above 0.99 for $\dttr\lesssim35$, i.e.\ despite 
missing rise time data (and the time delay 
estimation is insignificantly affected in this regime). Further truncation reduces 
the accuracy, in particular the ability of the neural network to distinguish unlensed and 2-image partial lightcurves. 

These results can be understood by 
considering that a typical supernova intrinsic rise 
time is $\sim20$ days, and with a single time delay 
of $\sim 12$ days for a 2-image system, then 
truncation of the first $\sim32$ days means one 
has only data beginning near maximum flux (and extending 
until the supernova fades substantially, more 
than a month after maximum). For 
4-image systems with sequential image time delays 
of $\sim 12$ days, while one might lose the first 
image for extended truncations, there is still 
substantial flux from at least 3 images to 
later times, and such a broad lightcurve would look 
quite different from the separated maxima of a long time delay 2-image 
system, so the accuracy remains high.

\section{Conclusions} \label{sec:concl} 

Strongly gravitationally lensed transients carry significant cosmological 
information, and will increasingly be discovered by upcoming surveys. 
While the use of well separated, resolved, multiple images and their 
fluxes is established, many instances will have individual images 
unresolved and their lightcurves blended together. We build on previous 
work on unresolved lensed transients by using deep learning to avoid 
restrictive assumptions about the lightcurve shape. 

Taking Type Ia supernovae as an example, we show that deep learning 
can classify the number of images with perfect precision and recall 
for time delays $\dt\gtrsim6$ days, accurately measure the individual 
time delays (to $\sigdt\approx1$ days) without significant bias, and 
do so $\sim1000\times$ 
faster than previous Monte Carlo fitting techniques. A simple 
combination of a classification neural net CNNc and a time delay 
measurement neural net is found to be highly efficient and accurate. 

We extended our analysis to variations of lens system and observational 
properties. The deep learning implementation is robust for noise up 
to $\sim20\%$ of maximum flux, has $\sim99\%$ accuracy even when 
rise time data is missing up to about maximum light, and still 
achieves $\sim95\%$ accuracy down to time delays $\dt\sim2$ days. 

While we have focused on Type Ia supernovae, the approach is generally 
applicable and can be tested for other transients in future work, 
including recognition of a transient as a Type Ia supernova. A major 
use of identification and characterization of unresolved transients 
is to engage followup resources for cosmological probes; for example, 
time delay cosmography will need not only time delay estimation 
but measurement of the lens galaxy mass profile or velocity dispersion. 
By identifying a set of promising lensed supernovae through our 
deep learning work, this can make the followup more efficient. 
Further aspects to be studied include the impact of microlensing, 
and different intrinsic lightcurve shapes, however the speed and 
accuracy of this deep learning approach should make those investigations 
more tractable than previous Monte Carlo fitting techniques.

\section*{Acknowledgements} 

We thank Satadru Bag, Alex Kim, and Arman Shafieloo 
for useful discussions. This work was supported in part 
by the Energetic Cosmos Laboratory. 
EL is supported in part by the U.S.\ Department of Energy, Office of Science, Office of High Energy Physics, under contract no.\ DE-AC02-05CH11231.

\section*{Data Availability}
The training and test data sets for simulated  lightcurves  of  1,  2,  and  4  image  systems in Sec.~\ref{sec:method} are available on the GitHub repository \href{https://github.com/mdeatecl/LensedSN124imagesLCs4DL}{https://github.com/mdeatecl/LensedSN124imagesLCs4DL}. 


\bibliographystyle{mnras}
\bibliography{LensedSNwDL}

\begin{thebibliography}{}
\makeatletter
\relax
\def\mn@urlcharsother{\let\do\@makeother \do\$\do\&\do\#\do\^\do\_\do\%\do\~}
\def\mn@doi{\begingroup\mn@urlcharsother \@ifnextchar [ {\mn@doi@}
  {\mn@doi@[]}}
\def\mn@doi@[#1]#2{\def\@tempa{#1}\ifx\@tempa\@empty \href
  {http://dx.doi.org/#2} {doi:#2}\else \href {http://dx.doi.org/#2} {#1}\fi
  \endgroup}
\def\mn@eprint#1#2{\mn@eprint@#1:#2::\@nil}
\def\mn@eprint@arXiv#1{\href {http://arxiv.org/abs/#1} {{\tt arXiv:#1}}}
\def\mn@eprint@dblp#1{\href {http://dblp.uni-trier.de/rec/bibtex/#1.xml}
  {dblp:#1}}
\def\mn@eprint@#1:#2:#3:#4\@nil{\def\@tempa {#1}\def\@tempb {#2}\def\@tempc
  {#3}\ifx \@tempc \@empty \let \@tempc \@tempb \let \@tempb \@tempa \fi \ifx
  \@tempb \@empty \def\@tempb {arXiv}\fi \@ifundefined
  {mn@eprint@\@tempb}{\@tempb:\@tempc}{\expandafter \expandafter \csname
  mn@eprint@\@tempb\endcsname \expandafter{\@tempc}}}

\bibitem[\protect\citeauthoryear{Bag, Kim, Linder  \& Shafieloo}{Bag
  et~al.}{2021}]{paper1}
Bag S.,  Kim A.~G.,  Linder E.~V.,   Shafieloo A.,  2021, \mn@doi [Astrophys.
  J.] {10.3847/1538-4357/abe238}, 910, 65

\bibitem[\protect\citeauthoryear{Bag, Shafieloo, Liao  \& Treu}{Bag
  et~al.}{2022}]{2110.15315}
Bag S.,  Shafieloo A.,  Liao K.,   Treu T.,  2022, \mn@doi [The Astrophysical
  Journal] {10.3847/1538-4357/ac51cb}, 927, 191

\bibitem[\protect\citeauthoryear{Barbary et~al.}{Barbary
  et~al.}{2020}]{sncosmo}
Barbary K.,  et~al., 2020, SNCosmo: Python library for supernova cosmology.
  Package version 2.1, \url {https://github.com/sncosmo/sncosmo}

\bibitem[\protect\citeauthoryear{{Biggio}, {Domi}, {Tosi}, {Vernardos},
  {Ricci}, {Paganin}  \& {Bracco}}{{Biggio} et~al.}{2021}]{2110.01012}
{Biggio} L.,  {Domi} A.,  {Tosi} S.,  {Vernardos} G.,  {Ricci} D.,  {Paganin}
  L.,   {Bracco} G.,  2021, preprint (\mn@eprint {arXiv} {2110.01012})

\bibitem[\protect\citeauthoryear{Boone}{Boone}{2021}]{parsnip}
Boone K.,  2021, \mn@doi [The Astronomical Journal] {10.3847/1538-3881/ac2a2d},
  162, 275

\bibitem[\protect\citeauthoryear{Davison, Parkinson  \& Tucker}{Davison
  et~al.}{2022}]{Davison_2022}
Davison W.,  Parkinson D.,   Tucker B.~E.,  2022, \mn@doi [The Astrophysical
  Journal] {10.3847/1538-4357/ac3422}, 925, 186

\bibitem[\protect\citeauthoryear{Denissenya, Bag, Kim, Linder  \&
  Shafieloo}{Denissenya et~al.}{2022}]{paper2}
Denissenya M.,  Bag S.,  Kim A.~G.,  Linder E.~V.,   Shafieloo A.,  2022,
  \mn@doi [Monthly Notices of the Royal Astronomical Society]
  {10.1093/mnras/stac143}, 511, 1210

\bibitem[\protect\citeauthoryear{Goldstein, Nugent  \& Goobar}{Goldstein
  et~al.}{2019}]{goldstein}
Goldstein D.~A.,  Nugent P.~E.,   Goobar A.,  2019, \mn@doi [Astrophys. J.
  Suppl.] {10.3847/1538-4365/ab1fe0}, 243, 6

\bibitem[\protect\citeauthoryear{Hsiao, Conley, Howell, Sullivan, Pritchet,
  Carlberg, Nugent  \& Phillips}{Hsiao et~al.}{2007}]{hsiao}
Hsiao E.~Y.,  Conley A.~J.,  Howell D.~A.,  Sullivan M.,  Pritchet C.~J.,
  Carlberg R.~G.,  Nugent P.~E.,   Phillips M.~M.,  2007, \mn@doi [Astrophys.
  J.] {10.1086/518232}, 663, 1187

\bibitem[\protect\citeauthoryear{{Huber} et~al.,}{{Huber}
  et~al.}{2019}]{1903.00510}
{Huber} S.,  et~al., 2019, \mn@doi [A\&A] {10.1051/0004-6361/201935370}, 631,
  A161

\bibitem[\protect\citeauthoryear{{Huber} et~al.,}{{Huber}
  et~al.}{2022}]{2108.02789}
{Huber} S.,  et~al., 2022, \mn@doi [A\&A] {10.1051/0004-6361/202141956}, 658,
  A157

\bibitem[\protect\citeauthoryear{Kingma \& Ba}{Kingma \&
  Ba}{2017}]{kingma2017adam}
Kingma D.~P.,  Ba J.,  2017, preprint (\mn@eprint {arXiv} {1412.6980})

\bibitem[\protect\citeauthoryear{Lochner et~al.,}{Lochner
  et~al.}{2022}]{2104.05676}
Lochner M.,  et~al., 2022, \mn@doi [The Astrophysical Journal Supplement
  Series] {10.3847/1538-4365/ac5033}, 259, 58

\bibitem[\protect\citeauthoryear{Muthukrishna, Parkinson  \&
  Tucker}{Muthukrishna et~al.}{2019}]{Muthukrishna_2019}
Muthukrishna D.,  Parkinson D.,   Tucker B.~E.,  2019, \mn@doi [The
  Astrophysical Journal] {10.3847/1538-4357/ab48f4}, 885, 85

\bibitem[\protect\citeauthoryear{Paszke et~al.,}{Paszke et~al.}{2019}]{pytorch}
Paszke A.,  et~al., 2019, in Wallach H.,  Larochelle H.,  Beygelzimer A.,
  d\textquotesingle Alch\'{e}-Buc F.,  Fox E.,   Garnett R.,  eds, , Advances
  in Neural Information Processing Systems 32.
Curran Associates, Inc., pp 8024--8035

\bibitem[\protect\citeauthoryear{Pierel, Rodney, Vernardos, Oguri, Kessler  \&
  Anguita}{Pierel et~al.}{2021}]{2010.12399}
Pierel J. D.~R.,  Rodney S.,  Vernardos G.,  Oguri M.,  Kessler R.,   Anguita
  T.,  2021, \mn@doi [Astrophys. J.] {10.3847/1538-4357/abd8d3}, 908, 190

\bibitem[\protect\citeauthoryear{Shu, Belokurov  \& Evans}{Shu
  et~al.}{2021}]{2011.04667}
Shu Y.,  Belokurov V.,   Evans N.~W.,  2021, \mn@doi [Mon. Not. Roy. Astron.
  Soc.] {10.1093/mnras/stab241}, 502, 2912

\bibitem[\protect\citeauthoryear{Springer \& Ofek}{Springer \&
  Ofek}{2021a}]{2101.11017}
Springer O.~M.,  Ofek E.~O.,  2021a, \mn@doi [Mon. Not. Roy. Astron. Soc.]
  {10.1093/mnras/stab1600}, 506, 864

\bibitem[\protect\citeauthoryear{Springer \& Ofek}{Springer \&
  Ofek}{2021b}]{2101.11024}
Springer O.~M.,  Ofek E.~O.,  2021b, \mn@doi [Mon. Not. Roy. Astron. Soc.]
  {10.1093/mnras/stab2432}, 508, 3166

\bibitem[\protect\citeauthoryear{{Verma}, {Collett}, {Smith}, {Strong Lensing
  Science Collaboration,}  \& {the DESC Strong Lensing Science Working
  Group}}{{Verma} et~al.}{2019}]{1902.05141}
{Verma} A.,  {Collett} T.,  {Smith} G.~P.,  {Strong Lensing Science
  Collaboration,}  {the DESC Strong Lensing Science Working Group} 2019,
  preprint (\mn@eprint {arXiv} {1902.05141})

\makeatother
\end{thebibliography}





\bsp	
\label{lastpage}
\end{document}